\documentclass[twocolumn,showpacs,preprintnumbers,amsmath,amssymb,superscriptaddress, bibnotes]{revtex4}

\usepackage{color}
\usepackage{graphicx}
\usepackage{dcolumn}
\usepackage{bm}
\usepackage{mathrsfs}
\begin{document}


\title{Spin Dynamics in Iron-based Layered Superconductor (La$_{0.87}$Ca$_{0.13}$)FePO\\ 
Revealed by $^{31}$P and $^{139}$La NMR Studies}

\author{Yusuke Nakai}
\email{nakai@scphys.kyoto-u.ac.jp}
\affiliation{Department of Physics, Graduate School of Science, Kyoto University, Kyoto 606-8502, Japan}
\author{Kenji Ishida}
\affiliation{Department of Physics, Graduate School of Science, Kyoto University, Kyoto 606-8502, Japan}
\author{Yoichi Kamihara}
\affiliation{ERATO-SORST, JST, Frontier Collaborative Research Center, Tokyo Institute of Technology, Yokohama 226-8503, Japan}

\author{Masahiro Hirano}
\affiliation{ERATO-SORST, JST, Frontier Collaborative Research Center, Tokyo Institute of Technology, Yokohama 226-8503, Japan}
\affiliation{Frontier Research Center, Tokyo Institute of Technology, Yokohama 226-8503, Japan}

\author{Hideo Hosono}
\affiliation{ERATO-SORST, JST, Frontier Collaborative Research Center, Tokyo Institute of Technology, Yokohama 226-8503, Japan}
\affiliation{Frontier Research Center, Tokyo Institute of Technology, Yokohama 226-8503, Japan}
\affiliation{Materials and Structures Laboratory, Tokyo Institute of Technology, Yokohama 226-8503, Japan}

\date{\today}

\begin{abstract}
We report $^{31}$P and $^{139}$La NMR studies of (La$_{0.87}$Ca$_{0.13}$)FePO, which is a family member of the recently discovered superconductor LaFeAs(O$_{1-x}$F$_x$). 
In the normal state, Knight shift and $1/T_1T$ show that Fermi-liquid state with moderate ferromagnetic fluctuations emerges below 30 K. 
From $1/T_1T$ of $^{31}$P and $^{139}$La, quasi-two dimensional electronic structure is suggested, in which the FeP layer is more conductive than the LaO layer.
In the superconducting (SC) state, although a clear Meissner signal was observed, $1/T_1T$ increases below $T_c$, in contrast to a decrease of $1/T_1T$ due to the opening of a SC gap, suggesting that novel low-energy spin dynamics develop in the SC state. 
\end{abstract}

\maketitle

The recent discovery of high-$T_c$ superconductor LaFeAs(O$_{1-x}$F$_x$) ($T_c$=26 K at $x$=0.11) \cite{KamiharaFeAs} and related materials which include the highest $T_c$ upto 55 K in the case of SmFeAs(O$_{1-x}$F$_x$) \cite{ZhiRenSm} except in some cuprates have attracted a great deal of attention. 
The parent compound LaFeAsO has a layered crystal structure (tetragonal $P4/nmm$) with alternately stacked LaO and FeAs layers, where Fe atoms form a two-dimensional (2-D) square lattice (see Fig.~1). 
Similary as in cuprates, superconductivity emerges when carriers are doped into the parent compound LaFeAsO which does not exhibit superconductivity but antiferromagnetic (AFM) order \cite{KamiharaFeAs}. 
Band calculations suggest multiband Fermi surfaces with five  Fe related bands crossing the Fermi level \cite{Lebegue, Singh} in contrast to the cuprates which have only one Fermi surface. 

One of the family compound LaFePO have similar Fermi surfaces as LaFeAsO \cite{Lebegue,Singh}, but some striking differences exist. 
LaFePO is a nonmagnetic metal with one order of magnitude smaller value of resistivity than LaFeAsO which undergoes an AFM order at $\sim$140 K accompanied with a structural phase transition at $\sim$160 K \cite{Dong, Cruz, Nomura, NakaiJPSJ}. 
Furthermore, $T_c$ of LaFePO is $\sim$4 K which is much lower than LaFeAs(O$_{1-x}$F$_x$)\cite{KamiharaFeAs, KamiharaFeP}. 
Therefore, uncovering the normal and SC state of LaFePO is important for understanding the mechanism of the high-$T_c$ superconductivity in LaFeAs(O$_{1-x}$F$_x$). 
In this Letter, we focus on $^{31}$P and $^{139}$La NMR studies of the LaFePO system, in order to reveal the magnetic properties as well as the SC gap properties. A brief comparison of NMR results between the LaFePO and LaFeAsO systems is shown.

For the present NMR study, we used Ca 13\% doped LaFePO instead of undoped one due to a large SC volume fraction as well as higher $T_c$ \cite{Kamihara2ndLaFePO}.
Polycrystalline (La$_{0.87}$Ca$_{0.13}$)FePO was synthesized by solid-state reactions \cite{KamiharaFeP, Kamihara2ndLaFePO}. 
Almost all the X-ray diffraction peaks were assigned to those of (La$_{0.87}$Ca$_{0.13}$)FePO although there were weak peaks 
($\sim$3\% relative to the most intense diffration peak of (La$_{0.87}$Ca$_{0.13}$)FePO, correponding to $\sim$2\% in volume fraction) attributed to antiferromagnetic FeP as an impurity phase. 
We note that FeP gives very little contribution to susceptibility because of no anomaly around 120 K  (see Fig.~3)\cite{FePimpurity}. 
Electrical resistivity measurements via a dc four-probe technique show the onset $T_c$ of $\sim8$ K, a large residual resistivity ratio of RRR $\equiv \rho(290K)/\rho(8K)$= 20 and a small residual resistivity of $\rho$(8 K)= 67 $\mu\Omega$cm. 
dc susceptibility was measured using a Quantum Design Physical Properties Measurement System (PPMS) with a vibrating sample magnetometer (VSM) option. 
The SC volume fraction was estimated to be $\sim$100 \% from the slope of $M$-$H$ curve at 2 K in a magnetic field range from zero to $H_{c1}$ \cite{Kamihara2ndLaFePO}.
Figure~1 shows the temperature dependence of ac susceptibility ($\chi_{\rm ac}$), which is obtained by measuring the self-inductance of an {\it in situ} NMR coil in a field-cooled condition. Clear Meissner signals with broad transitions are observed, which might suggest an anisotropy of $H_{\rm c2}$ due to 2-D electronic structure. 
In zero field, the onset ($T_c^{\rm onset}$) and midpoint ($T_c^{\rm mid}$) of the SC transition are estimated to be 8.0 and 5.9 K, respectively. 
\begin{figure}[tb]
\begin{center}
\includegraphics[width=7cm]{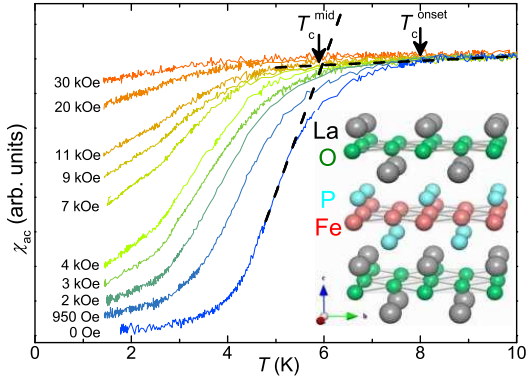}
\end{center}
\caption{(Color online) $T$-dependence of ac susceptibility obtained with an {\it in situ} NMR coil at $\sim$1.8 MHz. 
Inset: Crystal structure of LaFePO \cite{VESTA}.}
\label{ Meissner}
\end{figure}

Figure~2 shows $^{31}$P NMR spectra obtained by sweeping external field at a fixed frequency ($f=103.69$ MHz $\sim60$ kOe). The $^{31}$P NMR spectrum consists of a single and almost isotropic line as expected for an $I=1/2$ nuclei. Therefore, we determined the $^{31}$P Knight shift $^{31}K$ from the peak position of the spectrum. Note that $^{31}K=0$ was determined with using H$_3$PO$_4$ as a reference. 
\begin{figure}[tb]
\begin{center}
\includegraphics[width=6.5cm]{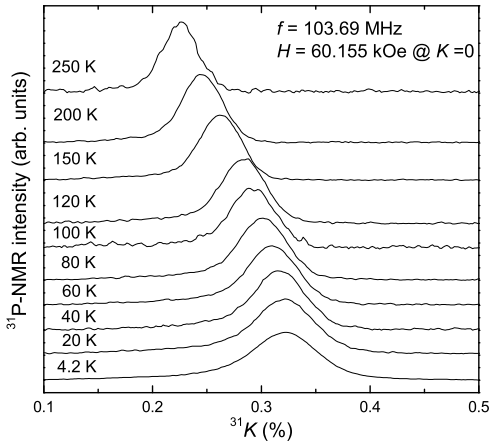}
\end{center}
\caption{Field-sweep $^{31}$P-NMR spectra at 103.69MHz.}
\label{ Spectra}
\end{figure}

Figure~3 shows the $T$-dependence of $^{31}K$, together with observed susceptibility $\chi_{\rm obs}$ (dotted line) and $\chi_{\rm obs}-\chi_{\rm imp}$ (solid line) which was obtained with subtracting impurity contributions from $\chi_{\rm obs}$ in 10 kOe. 
$\chi_{\rm obs}$ suggests that there are two dominant impurity contributions; one is from paramagnetic Ce$^{3+}$ ($\sim$0.13 mol\%) and another from ferromagnetic Fe$_2$P ($\sim$0.04 mol\%) which orders around 260 K \cite{Kamihara2ndLaFePO}.
$T$-dependence of the susceptibility originating from (La$_{0.87}$Ca$_{0.13}$)FePO, $\chi_{\rm obs}-\chi_{\rm imp}$, are extracted by subtracting the contributions of the paramagnetic Ce$^{3+}$ ion (giving a Curie-term at low $T$) and the ferromagnetic Fe$_2$P (yielding a step-like anomaly around 250 K). 
Although difference is seen below 50 K due to the Curie term, $^{31}K$ and $\chi_{\rm obs}-\chi_{\rm imp}$ exhibit a similar $T$-dependence above 60 K, yielding the hyperfine-coupling constant $^{31}A_{\rm hf} = 12.5 \pm 0.5$ kOe/$\mu_B$ and $T$-independent Knight shift $^{31}K_0=0.05\pm 0.01$ \% (see the inset of Fig.~3).
Since Knight shift represents intrinsic spin susceptibility of sample, the linear relation between $^{31}K$ and $\chi$ ensures that the subtracting procedure of the impurity contributions is valid at least between 60 K and 250 K. 
\begin{figure}[tb]
\begin{center}
\includegraphics[width=8cm]{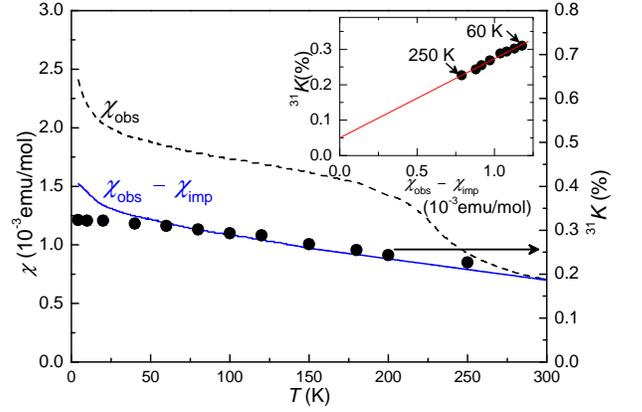}
\end{center}
\caption{(Color online) $T$-dependence of $\chi_{\rm obs}$ (dotted line), $\chi_{\rm obs}-\chi_{\rm imp}$ (solid line), and  the $^{31}$P Knight shift $^{31}K$ ($\bullet$). Error bars of $^{31}K$ are within the symbol size. Inset: $^{31}K$ vs ($\chi_{\rm obs}-\chi_{\rm imp}$) plot, giving $A_{\rm hf}$=12.5 kOe/$\mu_B$.} 
\label{}
\end{figure}

Figure~4 shows the $T$-dependence of $^{31}(1/T_1T)$ of $^{31}$P in $\sim$ 60 kOe. The spin-lattice relaxation rate $1/T_1$ was measured with a saturation recovery method and determined with a single component in the whole $T$-range (see Fig.~6). $^{31}(1/T_1T)$ increases with decreasing $T$, and saturates below 30 K, indicating that the normal state just above $T_c$ is in the Fermi-liquid state.
$T$-dependence of the resistivity exhibits $T^2$ behavior below 30 K, which is also characteristic of the Fermi-liquid state \cite{Kamihara2ndLaFePO}. 

In general, $1/T_1T$ is expressed with dynamical spin susceptibility $\chi''(\bm{q},\omega)$,
\begin{equation*}
\frac{1}{T_1T} = \frac{2\gamma_n^2~k_B}{(\gamma_e\hbar)^2}\lim_{\omega \rightarrow 0} \sum_{\bm{q}} |A(\bm{q})|^2 \frac{\chi''(\bm{q},\omega)}{\omega},
\end{equation*}
where $\gamma_e$ and $\gamma_n$ are the electronic and nuclear gyromagnetic ratios, respectively. $A(\bm{q}) = \sum_{j= 1 \sim 4} A_j\exp{(-i\bm{q}\cdot \bm{r}_j)}$ at the P site and $A_j$ is the hyperfine coupling constant between the P site and four nearest neighbor Fe electron spins and $\bm{r}_j$ is position vectors of Fe sites from the P site. $A(\bm{q})$ goes to zero at the wave vector $\bm{q} = (\pm\pi,\pm\pi)$. However, we note that $A(\bm{q})$ is not zero and $1/T_1T$ is enhanced if incommensurate AFM fluctuations with a wave vector $\bm{q}$ other than $(\pm\pi,\pm\pi)$ exist. 
Therefore, $1/T_1T$ of P is dominated by spin fluctuations in a broad $\bm{q}$ region around $\bm{q} = 0$.       

In a Fermi-liquid regime without strong magnetic correlations, the nature of spin fluctuations can be investigated from the spin part of Knight shift $K_{\rm spin}$ and $1/T_1T$ on the basis of the modified Korringa ratio $\mathscr{R} = (\gamma_e/\gamma_n)^2(\hbar/4\pi k_{\rm B})(T_1TK_{\rm spin}^2)^{-1}$; ferromagnetic (antiferromagnetic) correlations are suggested by the value $\mathscr{R}<1$ ($\mathscr{R}>1$).
Using the values of $^{31}(1/T_1T)$ and $^{31}K_{\rm spin}$ at 4.2 K, we obtain $\mathscr{R}=0.37$.
This relatively small $\mathscr{R}$ suggests that this compound has moderate ferromagnetic (FM) fluctuations, 
which is consistent with relatively large Wilson ratio $R_W\sim2.2-4.6$ \cite{KohamaFeP, HamlinSingleLaFePO, McQueenPolyLaFePO}. 
Although commensurate AFM fluctuations might be filtered at the P site, it is reasonable to consider that (La$_{0.87}$Ca$_{0.13}$)FePO possesses moderately FM correlations. 

In order to investigate the 2-D nature of the electronic structure, we performed $^{139}$La NMR in the LaO layer. 
Similar $T$-dependence of $1/T_1T$ of $^{139}$La as that of $^{31}$P is observed (see the inset of Fig.~4), suggesting that the LaO layer is also conductive as in the FeP layer.
It should be noted, however, that the value of $^{139}(1/T_1T) / ^{139}\gamma_n^2$ is one order of magnitude smaller than that of $^{31}(1/T_1T)/ ^{31}\gamma_n^2$ (see the each axis in the inset). 
Because $1/T_1T$ is a measure of the density of state at the Fermi energy in the Fermi liquid state, the large difference of $(1/T_1T)/\gamma^2_n$ of $^{31}$P and $^{139}$La implies quasi 2-D nature of the electronic states; the FeP layer is more conductive than the LaO layer. 
It is expected that magnetic fluctuations are affected by such 2-D electronic nature, which is related to $T$-dependence of $^{31}(1/T_1T)$ and $^{31}K$. SCR (Self-Consistent Renormalization) theory shows that $1/T_1T$ can be expressed as $1/T_1T \propto \chi(\bm{q} = 0)^{3/2 (1)}$ in the case when 2-D FM (3-D FM) fluctuations are dominant in the system \cite{Moriya}.
In the present case, $^{31}K_{\rm spin}^{3/2}$ shows a similar $T$-dependence with that of $^{31}(1/T_1T)$ (Fig.~4), which also suggests the predominance of 2-D FM fluctuations.
Taking all the results into account, we consider that the normal state in (La$_{0.87}$Ca$_{0.13}$)FePO is dominated by moderate FM fluctuations with 2-D nature. 
Band structure calculations suggest that LaFePO has multiple Fermi surfaces with 2-D nature and possible nesting between these bands. Among the nesting, we consider that magnetic fluctuations originating from the nesting with wave vector $\bm{q}\sim0$ contribute to $1/T_1$ and $K$. The involved nesting is likely to occur between two electron and/or two hole Fermi surfaces which are close in the $\bm{q}$-space, suggesting that the multiband structure of this system is important for understanding the magnetic fluctuations in contrast to the case of the cuprates.

\begin{figure}[tb]
\begin{center}
\includegraphics[width=9.5cm]{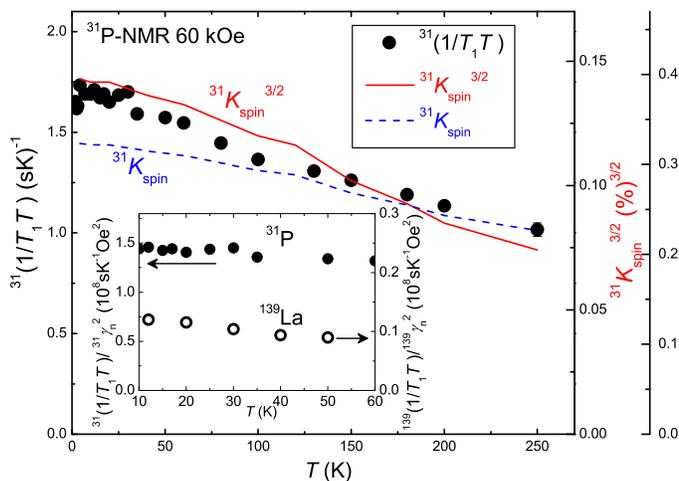}
\end{center}
\caption{(Color online) $T$-dependence of $^{31}(1/T_1T)$ of $^{31}$P in 60 kOe ($\bullet$) along with the spin part of the $^{31}$P Knight shift $^{31}K_{\rm spin}$ (dotted line) and $^{31}K_{\rm spin}^{3/2}$ (solid line). 
Inset: $(1/T_1T)/\gamma_n^2$ of $^{31}$P ($\bullet$) and $^{139}$La ($\circ$) as a function of temperature. The magnetic field for La NMR is 15.2 kOe.}
\label{invT1T}
\end{figure}
Let us now turn to NMR results in the SC state. 
The main panel of Fig.~5 shows $^{31}(1/T_1T)$ of $^{31}$P and $\chi_{\rm ac}$ in 950 Oe and 5.25 kOe, together with $^{31}(1/T_1T)$ in 60 kOe (normal state).
Although a clear Meissner signal is observed in $\chi_{\rm ac}$, $^{31}(1/T_1T)$ does not decrease, but starts to increase below $T_c$ followed by a continuous increase down to the lowest temperature. 
Similar increase of $1/T_1T$ of $^{139}$La below $T_c$ is also observed in 15.2 kOe. 
As shown in the inset of Fig.~5, While $^{31}(1/T_1)$ in 60 kOe crosses the origin within an experimental error (solid line), $^{31}(1/T_1)$ in 950 Oe appears to have a finite intercept. 
The increase of $1/T_1T$ below $T_c$ is apparent by showing the recovery of the nuclear magnetization $M(t)$ at a time $t$ after a saturation pulse. 
In Fig.~6, the recovery ($[M(\infty)-M(t)]/M(\infty)]$) in 950 Oe are plotted against $t\times T$ and the data are on the same curve when the values of $T_1T$ are the same. 
The data at 7 and 10 K are on the same curve, indicating that the Korringa relation holds in the $T$-region. However, the data at 1.45 K ($\textcolor{blue}{\blacksquare}$) show a steeper curve , indicating that $1/T_1T$ is larger than that in the normal state. 
The increase of $1/T_1T$ in the SC state is quite different from the decrease of $1/T_1T$ due to the opening of a SC gap observed in ordinary superconductors. 
Note that the increase is qualitatively different from a Hebel-Slichter coherence peak in $s$-wave superconductors, which shows a maximum around $T \sim 0.85 T_c$ \cite{Hebel, NakaiLaFe4P12}. 
From $^{31}$P NMR spectrum in 5.25 kOe, we found a broadening of the spectrum and its intensity decrease down to 60 \%  below $T_c$, which verify the occurrence of superconductivity. However, the broadening of the spectrum prevents us to determine Knight shift precisely below $T_c$ in this magnetic field.

\begin{figure}[tb]
\begin{center}
\includegraphics[width=8.5cm]{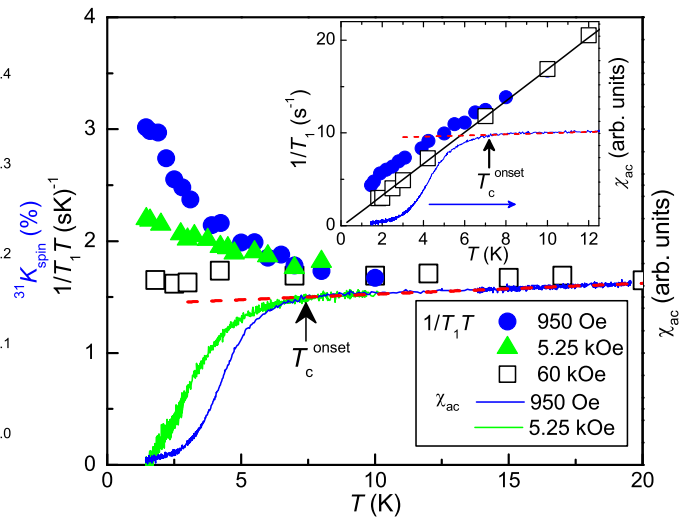}
\end{center}
\caption{(Color online) $T$-dependence of $^{31}(1/T_1T)$ and $\chi_{\rm ac}$ in 950 Oe ($\textcolor{blue}{\bullet}$ and blue solid line) and 5.25 kOe ($\textcolor{green}{\blacktriangle}$ and green solid line). 
Normal state $^{31}(1/T_1T)$ in 60 kOe is also plotted ($\square$). 
Inset: $^{31}(1/T_1)$ plot for 950 Oe and 60 kOe. The solid line represents a best fit of the $^{31}(1/T_1)$ below 20 K in 60 kOe. 
Broken lines in the main panel and the inset are linear fits to $\chi_{\rm ac}$ in the normal state, which define $T_c^{\rm onset}$.}
\label{}
\end{figure}
\begin{figure}[tb]
\begin{center}
\includegraphics[width=7cm]{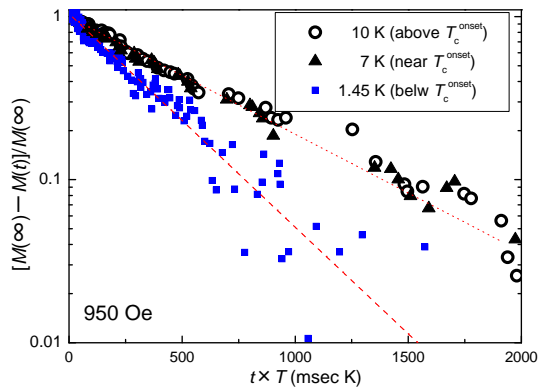}
\end{center}
\caption{(Color online) Relaxation curves vs $t\times T$ plot at 1.45 K, 7 K, and 10 K, which are fitted with a single exponential (dotted and broken lines).}
\label{}
\end{figure}
Next, we consider the origin of the low-energy spin dynamics below $T_c$. 
We discuss four possibilities which may account for the increase of $1/T_1T$. 
First, we consider impurity contributions detected by X-ray diffraction and susceptibility measurements. 
However, the $^{31}$P NMR peaks in Fig.~2 are free from such impurity phases, because no Curie term was observed in $^{31}K$ (see Fig.~3). 
In addition, if magnetic impurities were accidentally doped into the sample, it would be expected that the recovery curve of $M(t)$ should show multi-exponential behavior due to inhomogeneous relaxation rates. 
Obviously, this is not the case because the recovery curves show a single exponential behavior above and below $T_c$, as seen in Fig.~6. 
Therefore, we consider that it is unreasonable to ascribe the increase of $1/T_1T$ below $T_c$ to impurity contributions, although further studies using a cleaner sample is necessary to exclude the possibility. 
The second possibility is that vortex dynamics contribute to $1/T_1$ below $T_c$; the enhancement of $1/T_1T$ due to vortex dynamics have been sometimes observed in layered superconductors such as organic superconductors and high-$T_c$ cuprates \cite{DeSoto, Kanoda, Corti}. Vortex contributions to $1/T_1$ are especially significant at nuclear sites where quasiparticle DOS is small, {\it i. e.} $1/T_1T$ is small. 
In fact, $1/T_1T$ ascribed to the vortex dynamics have been detected in the very small $1/T_1$ ($1/T_1T\sim10^{-2}$-$10^{-3}$sec$^{-1}$K$^{-1}$ at the $^{1}$H site in $\kappa$-(ET)$_2$Cu[N(CN)$_2$]Br, and $\kappa-$(BEDT-TTF)$_2$Cu[N(CN)$_2$] \cite{DeSoto, Kanoda}, $1/T_1T\sim10^{-4}$sec$^{-1}$K$^{-1}$ at the $^{89}$Y site in YBa$_2$Cu$_4$O$_8$ \cite{Corti}). Since the increase of $^{31}(1/T_1T)$ is observed in $1/T_1T \sim 1.8$ sec$^{-1}$K$^{-1}$ which is $10^2$ times larger than $1/T_1T$ in such organic and cuprate superconductors, this possibility seems to be unlikely. 
The third possibility is the slowing down of the magnetic fluctuations due to the opening of a SC gap. 
Because the fluctuations enhanced by the scattering with conduction electrons are suppressed by the formation of the Cooper pairs, the characteristic energy of the fluctuations might decrease rapidly below $T_c$. However, as far as we know, such increase of $1/T_1T$ has not been reported in the SC state so far. 
The fourth possibility which includes our speculation is the emergence of low-energy fluctuations relevant to the SC order parameter. Especially, it has been suggested that the collective modes of the spin-triplet pairs give rise to novel spin dynamics in the SC state \cite{Vollhardt}. 
Although this possibility is very attractive, in order to check this possibility, $^{31}K$ measurements in the SC state by using a high-quality single crystal is desired. Further systematic $^{31}$P and $^{139}$La NMR investigations are needed to reveal the origin of the low-energy fluctuations which emerges below $T_c$, and are now in progress. 

Finally, we briefly compare the NMR results of (La$_{0.87}$Ca$_{0.13}$)FePO with those of LaFeAs(O$_{1-x}$F$_x$). We point out that the magnetic and SC properties of (La$_{0.87}$Ca$_{0.13}$)FePO are substantially different from LaFeAs(O$_{1-x}$F$_x$). Significant AFM fluctuations in the undoped and $x$=0.04 compounds and pseudogap behavior in $x$=0.11 are observed in LaFeAs(O$_{1-x}$F$_x$) with $^{75}$As NMR \cite{NakaiJPSJ}. In contrast, $^{31}$P NMR measurements suggest that (La$_{0.87}$Ca$_{0.13}$)FePO is a normal metal with moderate FM fluctuations. Moreover, we did not find a decrease of $1/T_1T$ below $T_c$ in (La$_{0.87}$Ca$_{0.13}$)FePO while a clear decrease of $1/T_1$ just below $T_c$ followed by $T^3$ dependence was observed in LaFeAs(O$_{1-x}$F$_x$). To gain a better understanding of these differences, it is necessary to study LaFeP(O$_{1-x}$F$_x$) using $^{31}$P and $^{139}$La NMR.

In summary, we report $^{31}$P and $^{139}$La NMR investigation of (La$_{0.87}$Ca$_{0.13}$)FePO. 
$^{31}(1/T_1T)$ and $^{31}K$ in the normal state suggest that the electronic state below 30 K is in the Fermi-liquid state with moderate ferromagnetic fluctuations. 
Quasi 2-D electronic structure is also inferred from the NMR measurements at the FeP and LaO layers. 
In the SC state, we observed an unusual increase of $1/T_1T$ below $T_c$, suggesting that novel low-energy spin dynamics develop in the SC state.

We thank Y. Kohama, Y. Maeno, S. Fujimoto, H. Ikeda and K. Yamada for valuable discussions.
This work was supported by the Grant-in-Aid for the 21st Century COE "Center for Diversity and Universality in Physics" from MEXT of Japan, and by the Grants-in-Aid for Scientific Research from JSPS.

\end{document}